\documentclass[10pt,preprintnumbers,showpacs,amsmath,amssymb,floatfix,twocolumn,prl]{revtex4}   
\usepackage{graphicx}
\usepackage{dcolumn}
\usepackage{bm}
\usepackage{longtable}
\usepackage{graphics}
\usepackage{amssymb}
\usepackage{amsmath}
\usepackage{xspace}
\usepackage{epsfig}
\newcommand {\dmit}{$\beta'$-dmit }
\newcommand {\dmitn}{$\beta'$-dmit}

\newcommand {\Br}{$\kappa$-Br }
\newcommand {\Brn}{$\kappa$-Br}
\newcommand {\Cl}{$\kappa$-Cl }
\newcommand {\NCS}{$\kappa$-NCS }

\newcommand {\tI}{$\theta$-I$_3$ }

\newcommand {\CN}{$\kappa$-CN$_3$ }
\newcommand {\CNn}{$\kappa$-CN$_3$}
\newcommand {\Cln}{$\kappa$-Cl}
\newcommand {\NCSn}{$\kappa$-NCS}

\newcommand {\tIn}{$\theta$-I$_3$}

\newcommand {\bdp}{$\beta''$-SO$_3$ }
\newcommand {\bdpn}{$\beta''$-SO$_3$}

\newcommand {\SROn}{Sr$_2$RuO$_4$}

\newcommand {\hlos}{$\frac12$LOS }
\newcommand {\hlosn}{$\frac12$LOS}
\newcommand {\qlos}{$\frac14$LOS }
\newcommand {\qlosn}{$\frac14$LOS}
\newcommand {\ibid}{{\it ibid}. }

\newcommand {\etalc}{{\it et al}., }
\begin{document}
\title{Mixed order parameters, accidental nodes and broken time reversal symmetry in organic superconductors: a group theoretical analysis}
\author{B. J. Powell}
\email{powell@physics.uq.edu.au} \affiliation{Department of Physics,
University of Queensland, Brisbane, Queensland 4072, Australia}


\begin{abstract}
We present a group theoretic analysis of several classes of organic
superconductors. We argue that highly frustrated half-filled layered
organic superconductors, such as $\kappa$-(ET)$_2$\-Cu$_2$\-(CN)$_3$
(where ET is BEDT-TTF) and $\beta'$-[Pd(dmit)$_2$]$_2X$, undergo two
superconducting phase transitions, the first from the normal state
to a $d$-wave superconductor and the second to a $d+id$ state. We
show that the monoclinic distortion of
$\kappa$-(ET)$_2$\-Cu\-(NCS)$_2$ means that the symmetry of its
superconducting order parameter is different from that of
orthorhombic $\kappa$-(ET)$_2$\-Cu\-[N(CN)$_2$]\-Br. We propose that
quarter filled layered organic superconductors, e.g.,
$\theta$-(ET)$_2$I$_3$, have $d_{xy}+s$ order parameters.
\end{abstract}

\maketitle

One of the most basic questions that can be asked about any phase of
matter is, what symmetries does it spontaneously break? For example,
all superconductors break gauge symmetry but many also break
additional symmetries \cite{James_adv_phys,Sigrist&Ueda}. Organic
charge transfer salts are an important class of superconductor
because they are highly tunable and have a number of exotic
properties \cite{Ross_review,Ishiguro,Kanoda,RVB_organics} such as a
small superfluid stiffness \cite{penetration,RVB_organics}, a Mott
transition \cite{Kagawa}, spin liquid states
\cite{Shimizu,CNrev,Kato}, a `bad metal' \cite{Ross_review},
pseudogap like behaviours \cite{nmr_review}, and charge ordered
states \cite{Merino}. Many of these properties are analogous to the
cuprates \cite{Ross_science} and this has heightened interest in
organic superconductors as model systems. Yet, in spite of the two
decades of intense effort to understand these systems, many basic
questions remain unanswered. Of particular importance is the, still
controversial, question of what is the symmetry of the
superconducting state \cite{disorder,Kuroki}.

In this Letter we use the methods of group theory to analyse the
symmetries of the superconducting states in these materials. We
discuss the pairing symmetries of
$\kappa$-(ET)$_2$\-Cu\-[N(CN)$_2$]\-Br (henceforth \Brn) and
$\kappa$-(ET)$_2$\-Cu\-[N(CN)$_2$]\-Cl (\Cln) which we argue are
`$d_{x^2-y^2}$', and $\kappa$-(ET)$_2$\-Cu\-(NCS)$_2$ (\NCSn) which
we show has additional components of other `$d$-wave' states and
thus accidental nodes (i.e., nodes that are \emph{not} required by
symmetry). Extending these arguments allows us to show that highly
frustrated materials such as $\kappa$-(ET)$_2$\-Cu$_2$\-(CN)$_3$
(\CNn) and $\beta'$-[Pd(dmit)$_2$]$_2X$ (\dmitn) \cite{dmit-foot}
will undergo two superconducting transitions the first from a normal
metal to a `$d$-wave' and the second from a `$d$-wave'
superconductor to a `$d+id$' state which breaks time reversal
symmetry (TRS). Similar reasoning implies that the quarter filled
organic superconductors have `$d_{xy}+s$' order parameters.

Group theory provides a powerful tool for addressing the symmetries
of superconducting states as it does not assume any particular
microscopic mechanism or theory of superconductivity
\cite{James_adv_phys,Sigrist&Ueda}. Such approaches are vital for
the organic superconductors where there are clear signs that BCS
theory, and indeed weak coupling approaches in general, are not
sufficient to explain the observed experimental results,
particularly the small superfluid stiffness
\cite{penetration,RVB_organics}. Further, the properties of nodal
quasiparticles, which are determined by the symmetry of the
superconducting state, have proved crucial in both the
superconducting and pseudogap states of the cuprates \cite{ARPES}.

Models of superconductivity based on both electron-phonon coupling
\cite{Kuroki} and the Hubbard model on an anisotropic triangular
lattice \cite{Ross_review,Kuroki} have been proposed for the
half-filled layered organic superconductors (\hlosn). Calculations
based on these two microscopic models lead to different conclusions.
For phononic mechanisms both isotropic, nodeless `$s$-wave' states
belonging to the $A_{1}$ irreducible representation (irrep) of
$C_{2v}$ and `$d_{x^2-y^2}$' states belonging to $B_{2}$ with nodes
in the gap along the lines $k_x=k_y$ and $k_x=-k_y$ have been
predicted. Calculations based on the Hubbard model suggest a ($B_2$)
`$d_{x^2-y^2}$' state \cite{RVB_organics,otherPRLs,Kuroki}.

\begin{table*}
\begin{tabular}{cccccccc}
  \hline
  Irrep & Required nodes
& Example basis functions ($k_z\|b$)
  & States &
    Example basis
    functions ($k_z\|c$)
  & States \\
  \hline
  $A_{1g}$ &  none & $1_{\bf k}$, $A_{\bf k}^2$, $B_{\bf k}^2$, $C_{\bf k}^2$, $X_{\bf k}Y_{\bf k}$ & $s$, $d_{xy}$ &  $1_{\bf k}$, $X_{\bf k}Y_{\bf k}$, $1_{\bf k}+X_{\bf k}Y_{\bf k}$  & $s$, $d_{xy}$\\
  $B_{1g}$ & line  & $A_{\bf k}B_{\bf k}$, $(X_{\bf k}+Y_{\bf k})Z_{\bf k}$ & $d_{(x+y)z}$ & $A_{\bf k}B_{\bf k}$, $X_{\bf k}^2-Y_{\bf k}^2$ & $d_{x^2-y^2}$ \\
  $B_{2g}$ & line & $A_{\bf k}C_{\bf k}$, $X_{\bf k}^2-Y_{\bf k}^2$  & $d_{x^2-y^2}$ & $A_{\bf k}C_{\bf k}$, $(X_{\bf k}+Y_{\bf k})Z_{\bf k}$ & $d_{(x+y)z}$ \\
  $B_{3g}$ & line & $B_{\bf k}C_{\bf k}$, $(X_{\bf k}-Y_{\bf k})Z_{\bf k}$ & $d_{(x-y)z}$ & $B_{\bf k}C_{\bf k}$, $(X_{\bf k}-Y_{\bf k})Z_{\bf k}$  & $d_{(x-y)z}$ \\
  \hline
\end{tabular}
\caption{The symmetry required nodes of the even parity irreps of
the point group $D_{2h}$, which represents the symmetry of the
orthorhombic organic superconductors such as \Brn, in which the
highly conducting plane is the $a$-$c$ plane, i.e., $k_z\|b$, and
\tIn, in which the highly conducting plane is the $a$-$b$ plane,
i.e., $k_z\|c$. The functions $1_{\bf k}$, $X_{\bf k}$, $Y_{\bf k}$,
$Z_{\bf k}$, $A_{\bf k}$, $B_{\bf k}$ and $C_{\bf k}$ may be any
functions which transform, respectively, as 1, $k_x$, $k_y$, $k_z$,
$k_a$, $k_b$ and $k_c$ under the operations of the group and satisfy
translational symmetry.}\label{tab:D2h}
\end{table*}

Experimentally, the pairing symmetries of \Brn, \Cln, and \NCS are
unclear. All show signs of unconventional superconductivity: there
are no Hebel-Schlicter peaks \cite{nmr_review}, the thermodynamic
measurements performed to the lowest temperatures show a power law
temperature dependence \cite{Carrington} and the disorder strongly
suppresses the superconducting critical temperature
\cite{disorder,Analytis}. Triplet superconductivity can be ruled out
\cite{disorder} on the basis of measurements of the Knight shift
\cite{nmr_review} and upper critical field.

\emph{Orthorhombic \hlosn.} Both \Br and \Cl are \hlos with
orthorhombic ($D_{2h}$) crystal structures. There are four even
parity irreps of $D_{2h}$ (see table \ref{tab:D2h}) all of which are
one dimensional. Canonically, the highly conducting plane is the
$ac$ plane in both \Br and \Cln. Further, it is usual to define the
$x$- and $y$-axes as lying along the directions of the largest
inter-dimer hopping integrals, which lie along the diagonals of the
$ab$ plane so that ${\bf \hat x}=({\bf \hat b}+{\bf \hat c})/2$ and
${\bf \hat y}=({\bf \hat b}-{\bf \hat c})/2$.

$\frac12$LOS crystals are extremely anisotropic: the inter-plane
hoping integral is about three orders of magnitude smaller than that
in-plane \cite{Ishiguro}. The superconducting properties are also
extremely anisotropic \cite{Ishiguro}. 
Hence, superconductivity in which the order parameter transforms as
either the $B_{1g}$ or $B_{3g}$ irreps are extremely unlikely
\cite{James_adv_phys}. Therefore, a `$d_{x^2-y^2}$' state
transforming as the $B_{2g}$ irrep of $D_{2h}$ is most consistent
with the experimental and theoretical evidence.

\emph{Monoclinic \hlosn.} The best studied \hlos with a monoclinic
crystal structure is \NCSn. This material has a $C_{2h}$ point group
and the highly conducting plane is the $bc$ plane. Here, the $x$-
and $y$-axes are usually taken to be along the diagonals of the
$bc$-plane. The only non-identity irrep of $C_{2h}$ that corresponds
to singlet superconductivity is $B_g$ (see table \ref{tab:C2h}).
This has symmetry required nodes only along the $c$-axis.

\begin{figure}
    \centering
    \epsfig{figure=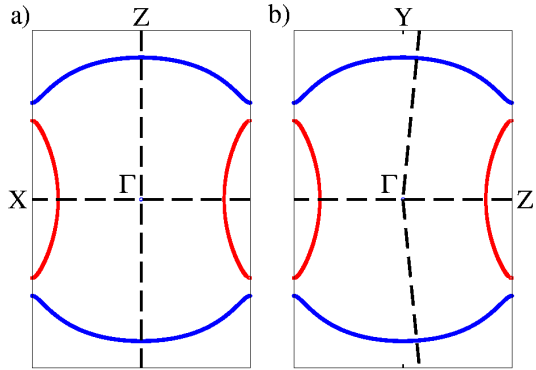, width=7cm, angle=0}
    \caption{(Color online.) Sketch of the differences between the nodal structures
    of (a) orthorhombic \hlos (e.g., \Brn) and (b) monoclinic \hlos
    (e.g., \NCSn). In both cases the Fermi surfaces are shown as solid
    lines and the nodes in the gap are shown as dashed lines.
    For $d_{x^2-y^2}$ states is orthorhombic \hlos symmetry requires that the nodes lie
    along the $a$- and $c$-axes. Whereas for $d$-wave states in monoclinic \hlos the only required
    symmetry is that the order parameter be antisymmetric under rotation by $\pi$ about the $z$-axis.
    As a node
    is not required along the $b$-axis in the `$d$-wave' state of the monoclinic
    \hlos we expect contributions from basis functions that
    do not go to zero along the $b$-axis to move this node and possibly remove this node
    altogether.
    High symmetry points are given their canonical labels.}\label{fig:nodes}
\end{figure}

However, a subset of the possible choices for the basis function of
the $B_g$ irrep lead to `$d_{x^2-y^2}$' superconductivity (i.e.,
states with nodes along $k_x^2=k_y^2$), which we have just argued is
the superconducting state realised in the orthorhombic \hlosn.
Further, the properties of \NCS are so similar to those of \Br that
the differences between the two materials are often described as
`chemical pressure' \cite{Kanoda}. Therefore, one expects that the
superconducting states of the two materials are closely related. If
the node is shifted away from the $b$-axis in the higher symmetry,
orthorhombic, case (as sketched in Fig. \ref{fig:nodes}) then there
is a change in symmetry which is accompanied by a phase transition.
There is no requirement for such as phase transition in the
monoclinic case as both possible the order parameters shown in Fig.
\ref{fig:nodes} transform as the $B_g$ irrep of $C_{2h}$. Any finite
contribution to the superconducting order parameter from basis
functions which do not have nodes along $b$-axis will cause this
node to move or, for a sufficiently large contribution, disappear.
Thus, predictions or measurements which show that monoclinic \hlos
have nodes anywhere except along the $c$-axis are not robust.
Therefore one cannot reliably conclude that the `$d$-wave' ($B_g$)
state is `$d_{x^2-y^2}$'. If there is an accidental node then it
will not lie along $b$-axis and it may have either a temperature
dependence, a $k_z$ dependence or both.

The above argument also shows that calculations based on the
anisotropic triangular lattice (whose symmetry is represented by the
group $C_{2v}$) excludes components of the gap in a manner that is
not relevant to the monoclinic \hlosn. Thus weak interactions
excluded by these models will dramatically change the symmetry of
the gap. Even neglecting such interactions, at sufficiently low
temperatures at least small additional `$d_{(x-y)z}$' components to
the order parameter should be expected. Exotic states with more
nodes than are present in `$d_{x^2-y^2}$' states have been proposed
\cite{Schmalian}. Such nodes would almost certainly be lifted in
both monoclinic and orthorhombic \hlos as they are far from robust.

\emph{TRS breaking in \CN and \dmitn.} Much attention \cite{CNrev}
has been focused on \CN following the discovery that, in spite of
there existing well formed local moments, in the low pressure,
insulating phase, these moments do not order down to the lowest
temperatures studied ($20$~mK) \cite{Shimizu}. Both Huckel
calculations \cite{CN_Huckel} and fits of the susceptibility
calculated by series expansions \cite{Shimizu,Zheng} to that
observed experimentally suggest that the band structure of \CN is
that of the isotropic triangular lattice. Series expansions
\cite{Zheng,Kato} also show that the triangular lattice Heisenberg
model is a good approximation for the low pressure, insulating phase
of \dmitn.

The symmetry of the triangular, or more correctly hexagonal, lattice
is represented by the $C_{6v}$ point group. An interesting feature
of $C_{6v}$ is that it has two two-dimensional (2D) irreps. In a 2D
irrep the order parameter, $\Delta_{\bf k}$, is a linear combination
of the basis functions, $\Psi_{\bf k}^{1,2}$, of the irrep, i.e.,
$\Delta_{\bf k}=\eta_1\Psi_{\bf k}^1+\eta_2\Psi_{\bf k}^2$. Hence,
on the hexagonal lattice the Ginzburg-Landau free energy of order
parameters belonging to the 2D irreps is
\cite{Sigrist&Ueda,James_adv_phys}%
\begin{eqnarray}
F_s-F_n&=&\alpha(T-T_c)(|\eta_1|^2+|\eta_2|^2) +
\beta_1(|\eta_1|^2+|\eta_2|^2)^2 \notag\\&& +
\beta_2(\eta_1^*\eta_2-\eta_1\eta_2^*)^2.
\end{eqnarray}
The ground state solution, $\vec{\eta}=(\eta_1,\eta_2)$, is: (i)
$\vec{\eta}=(1,0)$ or (ii) $\vec{\eta}=(0,1)$ for $\beta_2>0$ (the
degeneracy is lifted by sixth order terms
\cite{Sigrist&Ueda,James_adv_phys}); (iii) $\vec{\eta}=(1,i)$ for
$\beta_2<0$ (this is the weak coupling solution). In the $E_2$ irrep
$\Psi_{\bf k}^1$ describes a `$d_{x^2-y^2}$' state and $\Psi_{\bf
k}^2$ describes a `$d_{xy}$' state. Thus the three solutions
correspond to (i) `$d_{x^2-y^2}$' superconductivity; (ii) `$d_{xy}$'
superconductivity; and (iii) `$d_{x^2-y^2}+id_{xy}$'
superconductivity. Thus state (i) is the same state as we have
discussed above for the orthorhombic \hlosn. However, a number of
studies \cite{d+id_tri} of the Hubbard model on the hexagonal
lattice suggest that state (iii) is realised.

\begin{table}
\begin{tabular}{cccc}
  \hline
  Irrep & Required nodes & Example basis functions & States \\
  \hline
  $A_{g}$ & none & $1_{\bf k}$, $A_{\bf k}^2$, $B_{\bf k}^2$, $C_{\bf k}^2$ & $s$, $d_{xy}$ \\
  $B_{g}$ & 
  line & $\left\{ \begin{tabular}{c}
                              $A_{\bf k}C_{\bf k}$, $(X_{\bf k}-Y_{\bf k})Z_{\bf k}$, \\
                              $B_{\bf k}C_{\bf k}$, $X_{\bf k}^2-Y_{\bf k}^2$
                            \end{tabular} \right.$
    & $\left\{ \begin{tabular}{c} $d_{x^2-y^2}$,\\ $d_{(x-y)z}$ \end{tabular} \right.$\\
  \hline
\end{tabular}\caption{The symmetry required nodes of the
even-parity irreps of the group $C_{2h}$ which represents the
symmetry of \NCS and several other charge transfer salts with
monoclinic unit cells. Note that the symmetry line node in the $B_g$
irrep is required to lie in the plane
$k_c=k_x-k_y=0$.}\label{tab:C2h}
\end{table}

\begin{figure}
    \centering
    \epsfig{figure=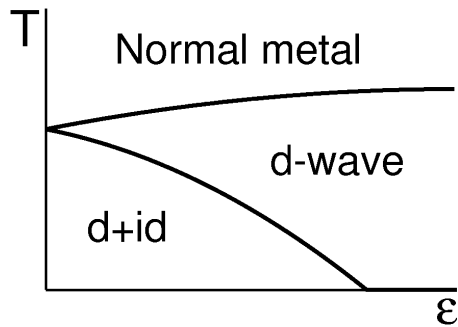, width=6cm, angle=0}
    \caption{Sketch of the proposed phase diagram for superconductivity
near the hexagonal lattice in \CN and \dmitn. $\varepsilon$ is a
symmetry breaking parameter which lowers the symmetry from $C_{6v}$
to $C_{2h}$. Physically $\varepsilon$ could represent uniaxial
strain or pressure. $\varepsilon\ne0$ at ambient pressure due to the
monoclinic crystal structure.}\label{fig:sketch}
\end{figure}

\emph{The role of monoclinicity.} It is important to realise that
although the band structures of \CN and \dmit are close to the
hexagonal lattice, they will not be precisely that of the hexagonal
lattice because these materials form monoclinic crystals. The
monoclinic distortion of the crystal lowers the symmetry of the
microscopic Hamiltonian. We account for this perturbation by
introducing a symmetry breaking field, $\varepsilon$ which, to
lowest order, enters the free energy as
$\varepsilon(|\eta_1|^2-|\eta_2|^2)$ \cite{Sauls}. Thus, 
\begin{eqnarray} F_s-F_n&=&\alpha_+|\eta_1|^2+\alpha_-|\eta_2|^2 +
\beta_1(|\eta_1|^2+|\eta_2|^2)^2 \notag\\&& +
\beta_2(\eta_1^*\eta_2-\eta_1\eta_2^*)^2,
\end{eqnarray}
where $\alpha_\pm=\alpha(T-T_{c\pm})$ and $T_{c\pm}=T_c\pm
\varepsilon/\alpha$. This theory predicts that the monoclinic
crystal will have two superconducting transitions, the first to
either a `$d_{x^2-y^2}$' superconducting state or a `$d_{xy}$'
superconducting state (both of which have nodes) and the second to
a, fully gapped, `$d_{x^2-y^2}+id_{xy}$' state. This leads us to
propose the phase diagram sketched in Fig. \ref{fig:sketch}. For
small $\varepsilon$ the difference in the two $T_c$'s grows linearly
with $\varepsilon$. But, for large $\varepsilon$ the symmetry of the
lattice will cease to be related to that of $C_{6v}$ and return to
that of $C_{2v}$. Therefore, for large $\varepsilon$ all of the
irreps are one-dimensional and, neglecting the possibility of
accidental degeneracies, we expect a single superconducting
transition. A similar scenario has previously been studied in some
detail in the context of the double superconducting transition
observed in UPt$_3$ \cite{Sauls}. However, for UPt$_3$ the proposed
symmetry breaking field arises from a weak antiferromagnetic
background \cite{Sigrist&Ueda}, rather than from the crystal
structure.

The above predictions are readily testable. Any number of
experiments might see the double superconducting transition (e.g.,
specific heat or ultrasound). Further, the proposed low temperature
`$d+id$' state breaks TRS, this could be detected directly by a
number of experiments \cite{Sigrist&Ueda} most notably $\mu$SR
\cite{Luke}. Experimental confirmation of a double superconducting
transition and broken TRS would be extremely important, not only
because of the intrinsic interest in these phenomena, but also
because they would be conclusive proof of unconventional
superconductivity in the layered organic superconductors. The
implication is then that phonons are not responsible for the
superconductivity in these materials, but rather strong electronic
correlations.

\emph{Quarter filled materials} have been less studied
experimentally, but, measurements of the in- and inter-plane
penetration depths show power law dependencies suggesting
unconventional superconductivity \cite{Giannetta}.  Some quarter
filled layered organic superconductors (\qlosn), e.g.,
$\theta$-(ET)$_2$\-I$_3$ (\tIn), have unit cells with a $D_{2h}$
point group, while others, e.g.,
$\beta''$-(ET)$_2$\-SF$_5$CH$_2$CF$_2$SO$_3$ (\bdpn), have $C_i$
point groups. It has been proposed \cite{Merino} that a minimal
model for \qlos is the quarter filled extended Hubbard (or
$t$-$U$-$V$-$J'$) model on the square lattice. (Note that the unit
cell used in this model is rotated with respect to the
crystallographic unit cell so that ${\bf\hat x}=({\bf\hat
a}+{\bf\hat b})/2$ and ${\bf\hat y}=({\bf\hat a}-{\bf\hat b})/2$.)
This has led to the idea that charge and spin fluctuations may
cooperatively mediate `$d_{xy}$' superconductivity \cite{Merino}.
However, this prediction of the location of the nodes is not robust
as the crystal lattices have significantly lower symmetries than the
model. In the $D_{2h}$ point group (table \ref{tab:D2h}) the basis
functions that describe the `$d_{xy}$' superconducting state belong
to the trivial $A_{1g}$ irrep. Therefore, these nodes are not robust
in the same sense as `$d_{x^2-y^2}$' states are not robust in
monoclinic \hlos. Thus one expects that the \qlos will have an
`$d_{xy}+s$' order parameter. The order parameter may still have
accidental nodes, but such nodes will be shifted away from the lines
$k_xk_y=0$. The same arguments and conclusions hold for \qlos with
$C_i$ point groups. A similar argument suggested that cuprates with
a small orthorhombic distortion are `$s+d_{x^2-y^2}$'
superconductors \cite{Sigrist}. Tunnelling experiments have shown
that the order parameter of YBCO does indeed have a significant
$s$-wave component \cite{Dynes}.

\begin{table}
\begin{tabular}{cccc}
  \hline
  Point group & Example material & Irrep & State \\
  \hline
  $C_{2h}$ ($C_{6v}$) & \CN & $B_g+iA_g$ ($E_2$) & `$d+id$' \\
  $D_{2h}$ & \Br & $B_{2g}$ & `$d_{x^2-y^2}$' \\
  $C_{2h}$ & \NCS & $B_{g}$ & `$d_{x^2-y^2}+d$' \\
  $D_{2h}$ & \tI & $A_{1g}$ & `$d_{xy}+s$' \\
  $C_i$ & \bdp & $A_{g}$ & `$d_{xy}+s$' \\
  \hline
\end{tabular}\caption{Summary of the superconducting states proposed
on the basis of the group theoretic analysis in this Letter. The
parenthetic point group and irrep in the first row indicate the
approximate symmetries which drive the physics.}\label{tab:summary}
\end{table}

In summary, we have presented a group theoretic analysis of the
several organic superconductors. This analysis has lead us to
propose order parameters summarised in table \ref{tab:summary}. In
detail we have argued that orthorhombic \hlos (e.g., \Br and \Cln)
have `$d_{x^2-y^2}$' order parameters but that monoclinic \hlos
(e.g., \NCSn) have nodes along the $c$-axis and possibly another
node in the conducting plane that is \emph{not} required by
symmetry. We proposed that highly frustrated \hlosn, such as \CN and
\dmitn, have `$d_{x^2-y^2}+id_{xy}$' order parameters but the
transition is split by the symmetry of the crystal so that there is
first a transition from the normal state to a `$d$-wave'
superconductor and then a second phase transition from the
`$d$-wave' superconductor to a `$d+id$' superconductor. We argued
that \qlosn, e.g., \tIn, have
`$d_{xy}+s$' order parameters. 

\acknowledgements

This work was motivated by conversations with J. Annett and R.
McKenzie. It is a pleasure to thank A. Ardavan, S. Blundell, B.
Braunecker, A. Carrington, A. Doherty, J. Fj{\ae}restad, M.
L\"uders, J. Merino, F. Pratt, and J. Varghese for useful
conversations and R. McKenzie for a critical reading of the
manuscript. I thank ISIS, and the Universities of Bristol and Oxford
for hospitality. This work was funded by the Australian Research
Council.

\end{document}